\documentclass[12pt]{iopart}
\usepackage{iopams}  
\usepackage{setstack}
\usepackage{graphicx}
\usepackage{bm}
\usepackage{array}
\begin{document}


\title[Quantum and classical simulation of CTC]{Closed timelike curves and chronology protection in quantum and classical simulators}

\author{Gonzalo Mart\'in-V\'azquez}
\address{Departamento de F\'isica Te\'orica, Universidad Complutense de Madrid, Plaza de Ciencias 1, 28040 Madrid,Spain
} \ead{gomart02@ucm.es}
\author{Carlos Sab\'in}
\address{Instituto de F\'isica Fundamental, CSIC,
Serrano, 113-bis,
28006 Madrid, Spain} \ead{csl@iff.csic.es}

\date{\today}

\begin{abstract}
In principe, General Relativity seems to allow the existence of closed timelike curves (CTC). However, when quantum effects are considered, it is likely that their existence is prevented by some kind of chronological protection mechanism, as Hawking conjectured. Confirming or refuting the conjecture would require a  full quantum theory of gravity. Meanwhile, the use of simulations could shed some light on this issue.  We propose simulations of CTCs in a quantum system as well as in a classical one. In the quantum simulation, some restrictions appear that are not present in the classical setup, which could be interpreted as an analogue of a chronology protection mechanism.
\end{abstract}

%
%
%
%
%

\section{\label{level1}Introduction}
It is well-known that in General Relativity there are, in principle, space-times where time travel is possible, that is, there are trajectories that form a loop over time, where an observer who follows them could return to its own past \cite{thorne}. These loops are called  Closed Timelike Curves (CTC). There is a close relationship between time travel and the possibility of achieving speeds larger than the speed of light in vacuum (superluminal velocity). Performing a path between two points at superluminal velocity and then the return path at a superluminal velocity in a different Lorentz frame, allows, in principle, to return to the origin before having even left  \cite{hawking} .

The existence of CTCs presents both logical problems (such as the well-known grandfather paradox) and theoretical ones \cite {lobo}. From the theoretical point of view, the presence of CTCs might be seen as an incompleteness of General Relativity itself: the evolution of a space-time with CTCs lacks a clear and consistent causal structure that can be described by General Relativity or other accepted theory. Certain conditions of realism not necessarily inherent to General Relativity (related to the type of matter (energy conditions) or to the asymptotic behavior of space-time, for example) are usually imposed to prevent the existence of CTCs and, thus, maintaining the causal structure \cite {thorne, curiel, mallary, hawking, wald}. It should be noted that the main interest in the study of CTCs lies in the search for physical mechanisms that prevent their creation \cite {thorne}, such as the chronology protection conjecture proposed by Hawking \cite {hawking}. In fact, the most promising route of research comes from the combination of the theory of General Relativity and quantum field theory in curved space-times, which could help to understand some aspects of quantum gravity \cite {thorne}. However, only a full theory of quantum gravity could finally close this open problem, by confirming or refuting Hawking's conjecture.

In physical problems of this nature, due to the difficulty (or impossibility) of observing the phenomenon itself, the use of simulations might be interesting, both in classical and quantum setups. Using classical means, processes such as superluminal motion \cite {clerici} or the formation of an event horizon in a white hole \cite {philbin} can be simulated. The use of quantum simulators has recently brought results in physical processes of difficult or dubious observation, such as the simulation of a traversable wormhole \cite {sabworm}, space-times in which superluminal trips are allowed and even CTCs \cite {sabexot}, Hawking radiation \cite {nation}, magnetic monopoles \cite {ray} or tachyonic particles \cite {lee}. The nature of each problem makes it necessary to use different types of systems to perform the simulations.

In this paper we are interested in the analysis of the possible mechanisms in charge of preventing the existence of CTCs. In the absence of a full theory of quantum gravity, an experimental simulator including quantum effects might shed light on this open issue. There are many proposed space-times that allow the existence of CTCs, each of them with more or less reasonable physical properties \cite {thorne}. We will focus only on the recent proposal by Mallary \textit {et al.} \cite {mallary}, a space-time consisting of a wire of matter of infinite length that can be moved at relativistic speeds, whose line element is given by
\begin{equation}
\ ds^{2}=-Fc^{2}dt^{2}+\frac{1}{F}dr^{2}+dz^{2}+r^{2}d\phi^{2},
\label{eq1}
\end{equation}
where
\begin{equation}
\ F=\cases{1+\left ( \frac{1}{r} -\frac{1}{R}\right )^{n} &if $r\leq R$ \\ 1 &if $ r> R$\\},
\label{eq2} 
\end{equation}
where the radius $ R $ is a positive arbitrary constant and $n \geq 2$. The term $ F $ ensures that the radius of the wire is finite and presents a singularity at $ r = 0 $ (which leads to an infinite mass per unit length). This metric violates the hypothesis of cosmic censorship, since it lacks an event horizon because the $ \frac {1} {F} $ factor of the radial coordinate never becomes infinite. It satisfies the  weak, null and strong energy conditions. However, it does not meet the dominant energy condition \cite {curiel}. Although similar metrics can be considered which fulfill the dominant energy condition as well as others having a finite size, the study of CTCs in these metrics is more involved and will not be addressed here \cite {mallary}.

We propose the experimental simulation of photon paths in the space time described by the metric (\ref {eq1}) that give rise to CTCs. For this we will consider two essentially different systems: a classical one and a quantum one. As a classical system we will consider the signal observed by the scattering of a light front on an inclined surface \cite {clerici} and as a quantum system we will use a superconducting circuit, specifically an \textit { array} of SQUIDs \cite {simoen, lahten, sabworm}. 
In both cases, simplified versions of (\ref {eq1}) are assumed, where the paths followed by the photons are carried out in a single spatial dimension, that is, they are constrained to a $1+ 1D$ section of the full spacetime. Then we can generally use:
\begin{equation}
\ ds^{2}=-c^{2}\left ( \rho ,t \right )dt^{2}+d\sigma ^{2},
\label{eq3}
\end{equation}
where $\rho$ and $\sigma$ are arbitrary coordinates (if $ \rho $ does not match $ \sigma $, it is taken as an additional parameter). In this way we have Minkowski-like space-times with an effective light speed that depends on spatio-temporal coordinates. Then, in order to implement the space-time (\ref {eq1}), a $1 + 1D$ section is taken from the full $3 + 1D$ space-time, so that two of the coordinates are ignored, obtaining a dimensionally reduced space-time, with an expression of the form (\ref{eq3}). For axial trajectories on the axis $z$ ($\rho = r $; $ \sigma = z $) the metric of the dimensionally reduced space-time is finally:
\begin{equation}
\ ds_{z}^{2}= -c_{v}^{2}Fdt^{2}+dz^{2}
\label{eq4} \\
\end{equation}
where $ c_{v} $ is the speed of light in vacuum. Then, we see in Eq. (\ref{eq4}) that we have $r$-dependent effective speed of light, $ c_{z}=c_{v}\sqrt{F} $.

The structure of the paper is the following. In Section \ref{level2} we will briefly describe the particular CTC proposed in \cite{mallary}. Then we will consider in Section \ref{level3} the quantum simulation of the spacetime where these CTCs arise and discuss the restrictions that appear when we try to implement the CTC. Finally, we will see in Section \ref{level4} that these mechanisms are absent in a classical setup, where we can actually propose a realistic analogue of a CTC. We conclude in Section \ref{level5} with a summary of our results.

\section{\label{level2}Existence and features of CTCs}

\begin{figure}[!]
\includegraphics[width=\columnwidth]{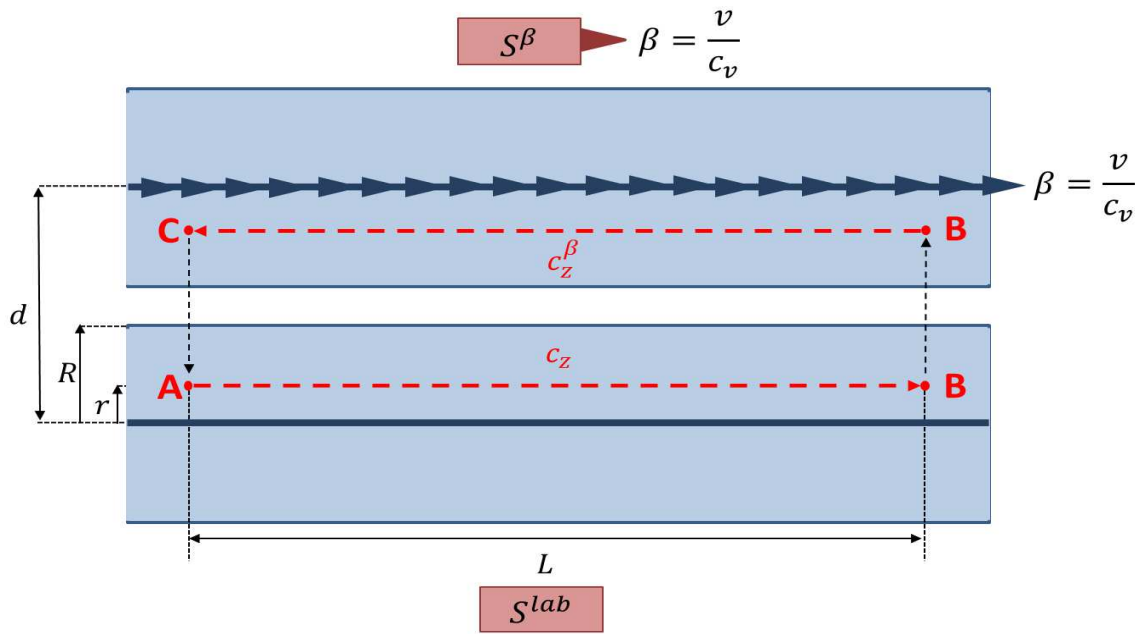}
\caption{\label{fig1} Diagram of the CTC in the space-time formed by two wires that move with a certain relative speed. $ S ^ {lab} $ represents the observing reference system that is at rest with the lower wire and $ S ^ {\beta} $ is a reference system at rest with respect to the upper wire that has a boost $ \beta $ relative to $ S ^ {lab} $. Both wires have an infinite length in the longitudinal axis.}
\end{figure}

Before trying to implement a simulation of CTCs, we will briefly summarize the necessary conditions so that in the space-time (\ref {eq1}) a CTC can be produced, following the more detailed description of \cite {mallary}. Two separate parallel wires will be needed at a distance $d$ ($2R <d \ll L $, where $L$ is a physical distance traveled on the axis $ z $), one of which moves at relativistic speed in the direction of the axis $ z $ as shown in Figure ~ \ref {fig1}. In this way, the wires do not interact gravitationally, and are separated by an empty space-time. Two reference systems at rest at great distances from both wires, denominated $ S ^ {Lab} $ and $ S ^ {\beta}$, will be considered for the lower and upper wire, respectively. Then the upper wire will undergo a  boost $ \beta = \frac {v} {c_ {v}} <1 $ in the direction of the $ z $ axis. A photon which follows the path described in Figure ~ \ref {fig1} travel through regions where its speed is greater than the speed of light in vacuum, which will depend on the distance from the wire, as it is deduced from (\ref {eq4}).
We assume that the photon makes a path in the lower wire (at rest with the laboratory) and the same path back in the upper wire in the opposite direction to the boost with a velocity $ c_ {z} = c_ {v} \sqrt {F} \geq c_ {v} $, where the radial distances traveled will be neglected (since $ d \ll L $).
It is shown in \cite{mallary}, that the total time for the round trip can be negative if:
\begin{equation}
\ \beta > \frac{2}{\sqrt{F}+\frac{1}{\sqrt{F}}}.
\label{eq17}
\end{equation} 
Thus, (\ref{eq17}) is the CTC condition as described in \cite{mallary} without the need to explicitly compute metric of the space-time with CTC (we refer to the original paper for further explanation). This scenario can be generalized for the case where in the first wire (at rest with respect to the laboratory system $ S ^ {Lab} $) the coordinate speed of light is $ c_ {v} \sqrt{F_ {1}} $ and for the second wire (at rest with respect to the system with boost $ S ^ {\beta} $) the speed of the coordinate light is $ c_ {v} \sqrt{F_ {2}}$. In this way, we find that the CTC condition takes the general form
\begin{equation}
\ \beta > \frac{1+\frac{\sqrt{F_{2}}}{\sqrt{F_{1}}}}{\frac{1}{\sqrt{F_{1}}}+\sqrt{F_{2}}}
\label{eq18}
\end{equation}
which is clearly reduced to (\ref{eq17}) when $ F_{1}=F_{2} $

\section{\label{level3}Quantum simulation of CTCs}

Our aim is to simulate the path described in Figure ~ \ref {fig1}, trying to reach the CTC condition (\ref{eq17}). For the quantum simulation of the space-time described by (\ref{eq4}) we consider the conformal invariance of the Klein-Gordon equation for a scalar field in $ 1+1D $ \cite {birrel}. Essentially, this is the case for an electromagnetic wave in an open transmission line with an array of dc-superconducting quantum interference devices (dc-SQUIDs) embedded on it \cite {lahten, simoen}. For our purposes, a SQUID can be considered as a tunable Josephson junction (JJ), namely a nonlinear inductance which can be controlled by an external magnetic flux. (For more details on the physics of JJs and SQUIDs in the context fo modern quantum technologies, see for instance \cite{simoen,reviewnori}). In such a way, the propagation speed of a microwave quantum electromagnetic field along the transmission line is given by $ c = \frac {1} {\sqrt {LC}} $, where $ L $ and $ C $ are the inductance and capacitance per unit length, respectively. Since the number of SQUIDs embedded in the transmission line is large enough, we can consider that $ L = L_ {s} $ and $ C = C_ {s} $, where $ L_ {s} $ and $ C_ {s} $ are the inductance and capacitance of a single SQUID. Note that, actually, the capacitance and inductance per unit length are $ \frac{C_{s}}{\epsilon} $ and $ \frac{L_{s}}{\epsilon} $, respectively, where $ \epsilon $ is the size of the SQUID; this does not affect the results since $ \epsilon^{2} $ get absorbed in the definition of $ c_{0} $ (see below). If the SQUID area is small enough, its self-inductance can be neglected. Each SQUID has two JJs but, considering that both have the same critical current ($ I_ {c} $), it can be treated as a single Josephson junction whose inductance is given by
\begin{equation}
\ L_{s}\left ( \phi _{ext} \right )= \frac{\phi _{0}}{4\pi I_{c}\cos \frac{\pi \phi _{ext}}{\phi _{0}}\cos \psi }
\label{eq30},
\end{equation}
where $\phi_ {0} = \frac {h} {2e} $ is the flux quantum, $\phi_ {ext} $ is the external magnetic threading the SQUID  and $ \psi $ is the phase difference along the SQUID, which we will take in the weak signal limit $\psi = 0$. In this way, the speed of light in the transmission line is
\begin{equation}
\ c^{2}\left ( \phi _{ext} \right )= \frac{1}{L_{s}C_{s}}= \frac{1}{\frac{\phi _{0}}{4\pi I_{c}}C_{s}}\cos \frac{\pi \phi _{ext}}{\phi _{0}}= c_{0}^{2}\cos \frac{\pi \phi _{ext}}{\phi _{0}}
\label{eq31}
\end{equation}
where $ c_ {0} = c (\phi_ {ext} = 0) = \frac {1} {\sqrt {L_ {s} (\phi_ {ext =0}) C_ {s}} }$ is the speed of light in the transmission line in the absence of external magnetic flux. To modify the velocity (\ref{eq31}) along the SQUID array, a magnetic flux $\phi_ {ext} $ will be applied, with a time and space dependence suitable to emulate the section of space-time of interest. First, we will divide this magnetic flux into two components, such that
\begin{equation}
\ \phi _{ext}\left ( r,t \right )= \phi _{ext}^{DC}+\phi _{ext}^{AC}\left ( r,t \right )
\label{eq32}
\end{equation}
As shown in \cite{sabexot}, we get:
\begin{equation}
\ c^{2}\left ( \phi _{ext} \right )=c^{2}\left ( \phi _{ext}^{DC} \right )\tilde{c}^{2}\left ( \phi _{ext} \right ),
\label{eq33}
\end{equation}
where
\begin{eqnarray}
\ c^{2}\left ( \phi _{ext}^{DC} \right )&=c_{0}^{2}\cos \frac{\pi \phi _{ext}^{DC}}{\phi_{0}}
\label{eq34} \\
\ \tilde{c}^{2}\left ( \phi _{ext}\right )&=\sec \frac{\pi \phi _{ext}^{DC}}{\phi_{0}}\cos \frac{\pi \phi _{ext}}{\phi _{0}},
\label{eq35}
\end{eqnarray}
under the restriction $ (\frac{\pi \phi_{ext}^{DC}}{\phi_{0}}) $,$ (\frac{\pi \phi_{ext}}{\phi_{0}}) \in \left [ -\frac{\pi}{2},\frac{\pi}{2} \right ] $. For the simulation of the spacetime in (\ref{eq4}), we first set an equivalence between the speed of light in vacuum $ c_{v}^{2} $ and $ c^{2}(\phi_{ext}^{DC}) $ such that
\begin{equation}
\ c_{v}^{2}\sim c_{0}^{2}\cos \frac{\pi \phi _{ext}^{DC}}{\phi _{0}}
\label{eq36}
\end{equation}

In this way, setting a constant magnetic flux $\phi_ {ext} ^ {DC}$ will simulate our simulated flat- spacetime speed of light, which might be significantly smaller than the actual value of the speed of light in vacuum $c_ {v}$ and that the speed of light in the transmission line in the absence of magnetic flux $c_ {0}$. In summary, we replace the actual speed of light by a virtual different one, and we assume that the latter plays the same role as the real speed of light but in a virtual universe, setting a virtual causal structure. The superluminal motion obtained is referred to this virtual speed of light, but it is always subluminal with respect to the real speed of light. In this way, we can build the analogue of a CTC with respect to the virtual speed of light but without sending any physical object back in time in any sense. This will be necessary to simulate superluminal velocities in the superconducting circuit.
Secondly, the $AC$ component of the magnetic flow $\phi_ {ext} ^ {AC} (r, t)$ will be used to simulate a spatiotemporal profile for the speed of light such that:
\begin{equation}
\ F=\sec \frac{\pi \phi _{ext}^{DC}}{\phi _{0}}\cos \frac{\pi \phi _{ext}}{\phi _{0}}
\label{eq37} \\
\end{equation}
Thus, we will need the following profiles for the magnetic fluxes:
\begin{equation}
\ \frac{\pi \phi _{ext}^{AC}\left ( r,t \right )}{\phi _{0}}=\arccos \left (F\cos \frac{\pi \phi _{ext}^{DC}}{\phi _{0}} \right )-\frac{\pi \phi _{ext}^{DC}}{\phi _{0}}
\label{eq39} \\
\end{equation}
Since the path takes place at a constant radial distance, the speed of the coordinate light will be identical and constant for each reference system. However, for the simulation, we can only simulate an effective speed of light for the laboratory system $ S ^ {lab} $. In the first case it is immediate, since the speed of the simulated light will simply be $c_ {z} = c_ {v} \sqrt {F}$. The magnitude of $ c_{z} $ will be limited by the magnetic flux value $ \phi_{ext}=\frac{\phi_{0}}{2} $; close values to this limit will cause quantum fluctuacions in the superconductor phase $ \psi $ due to the array impedance, invalidating the approximation made \cite {simoen}. Considering the simulable limit $ \phi_{ext}=0.45 \phi_{0} $ and (\ref{eq39}) we obtain an upper value of $ c_{z} \sim 2.5 c_{v} $ \cite {sabworm}. In the case of the path in the upper wire, we are not able to boost a transmission line up to relativistic speeds. Then, we directly state what would be the metric of the wire described by (\ref {eq1}) in motion along the direction of increasing $ z $ ($ z> 0 $) with a certain velocity $ v $ (where $ r $ is a parameter). That is, what would be the metric observed from rest in the distance (flat spacetime) of a moving wire.

\begin{figure}[!]
\includegraphics[width=\textwidth]{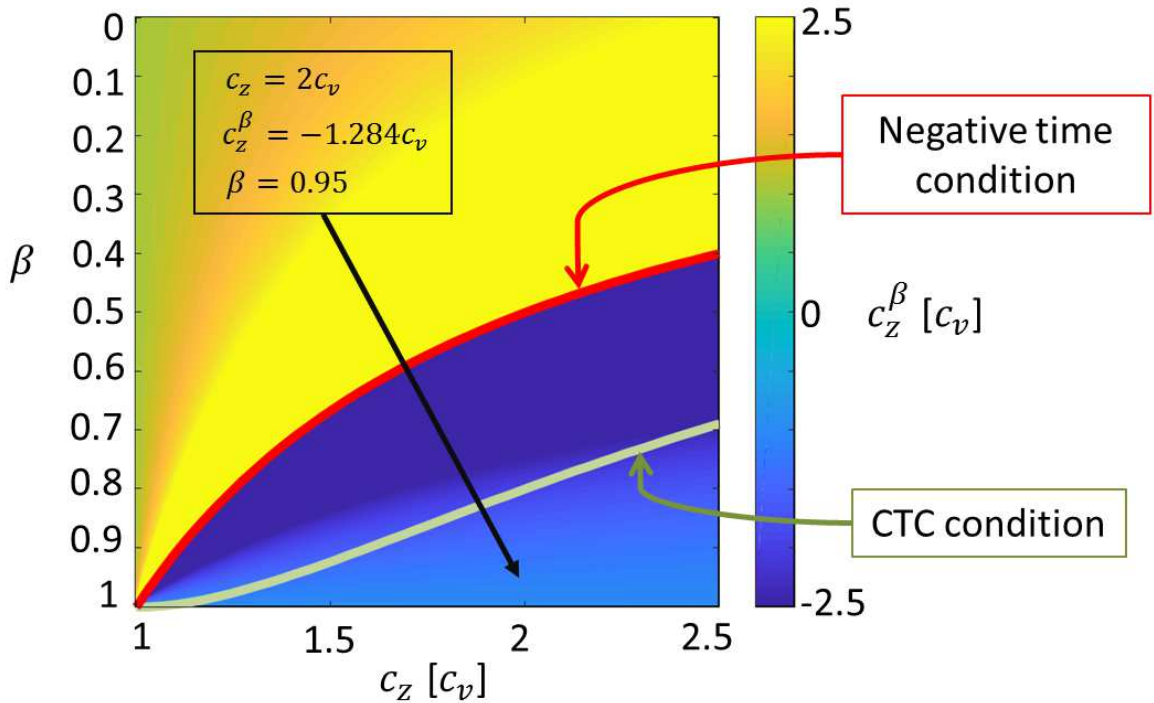}
\caption{\label{fig2} Simulated speed of light in the upper wire $c_{z}^{\beta}$ vs simulated $\beta$ and speed of light in the bottom wire $c_z$. The points under the red curve correspond to negative time for the $BC$ path in the upper wire in the laboratory coordinate system, while the points  fulfilling the CTC condition are under the light green line. In both cases, $c_{z}^{\beta}$ is negative and therefore out of experimental reach. The black arrow corresponds to a particular example.}
\end{figure}

We consider two reference systems, one $ S $ with coordinates ($ z, t $) static with respect to the wire in motion and another $ S '$ with coordinates ($ z', t '$) moving with the wire. The relations between the coordinates of both reference systems are given in the standard way:

\begin{eqnarray}
\ t'&= \gamma \left ( t-\frac{v}{c_{v}^{2}}z \right )
\label{eq43} \\
\ z'&=\gamma \left ( z-vt \right ),
\label{eq44}
\end{eqnarray}
where $ \gamma=1/\sqrt{1-\frac{v^{2}}{c_{v}^{2}}} $ is the usual Lorentz factor. Using this, we find:
\begin{equation}
\ ds^{2}=-\gamma ^{2}F\left ( c_{v}dt-\beta dz \right )^{2}+\gamma ^{2}\left ( dz-\beta c_{v}dt \right )^{2}.
\label{eq48}
\end{equation}
Eq. (\ref{eq48}) possesses two families of null geodesics: 
\begin{eqnarray}
\ dz&=c_{v}\frac{\beta +\sqrt{F}}{1+\sqrt{F}\beta }dt\underset{\beta =0,F=1}{\rightarrow}dz=c_{v}dt
\label{eq49} \\
\ dz&=c_{v}\frac{\beta -\sqrt{F}}{1-\sqrt{F}\beta }dt\underset{\beta =0,F=1}{\rightarrow}dz=-c_{v}dt,
\label{eq50}
\end{eqnarray}
where, by considering the flat-spacetime limit ($F = 1$) for a wire at rest ($\beta = 0$), we see that (\ref {eq49}) corresponds to the path of a photon in the same direction as the motion of the wire while (\ref {eq50}) corresponds to the opposite direction. In the latter case, we have that the speed of light is:
\begin{equation}
\ c_{z}^{\beta }=c_{v}\frac{\sqrt{F}-\beta }{1-\sqrt{F}\beta },
\label{eq41}
\end{equation}
which can be negative -- and then reverse the time direction, a necessary but not sufficient condition for a CTC-- if 
 \begin{equation}
 \sqrt {F} \beta> 1.\label{CTCcondition2}
 \end{equation}
We analyze the possibility of achieving the condition (\ref{CTCcondition2}) and the more restrictive (\ref{eq17})  in Figure ~ \ref {fig2}. For a coordinate speed of light in the wire at rest with the laboratory, a corresponding range of $\beta$ in the boosted wire would give rise to a CTC, which in turn translates into a range of values for $c_{z}^{\beta}$ that are compatible with a CTC. As can be seen in Figure ~ \ref {fig2}, $c_{z}^{\beta}$ is always negative in the CTC region. However, we are not able to simulate an effective negative speed of light by means of Eqs. (\ref{eq34}) and (\ref{eq35}). Therefore, a fundamental restriction appears in this quantum setup, preventing us from generating a CTC.

\begin{figure}[!]
\includegraphics[width=\columnwidth]{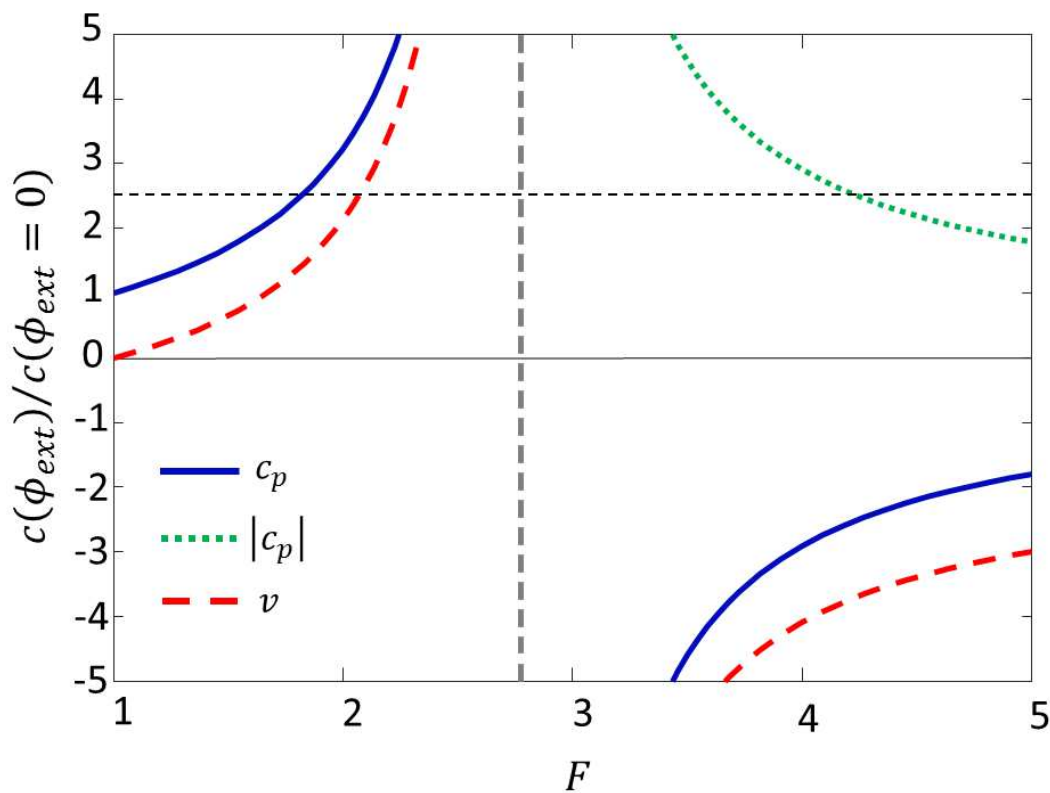}
\caption{\label{fig3} Light propagation and pulse velocities for the particular case $ \beta=0.6 $. The horizontal black dashed line corresponds approximately to maximum simulable value when reducing the background speed of light using $ DC $ magnetic fluxes; note that both $ c_{p} $ and $ v $ depend on $ c_{v} $, where $ c_{v}^{2}\sim c_{0}^{2}\cos \frac{\pi \phi _{ext}^{DC}}{\phi _{0}} $. The absolute value for $ c_{p} $ is represented only when $ c_{p} $ is negative. }
\end{figure}

Interestingly, defining: 
\begin{eqnarray}
\ c_{p}&=c_{v}\frac{\sqrt{F}\left ( 1-\beta ^{2} \right )}{1-F\beta ^{2}}
\label{eq54} \\
\ v&=c_{v}\frac{\beta \left ( F-1 \right )}{1-F\beta ^{2}}, 
\label{eq55}
\end{eqnarray}
the metric (\ref{eq48}) can be rewritten as:
\begin{equation}
\ ds^{2}=-\left ( c_{p}^{2}-v^{2} \right )dt^{2}+2vdtdz+dz^{2},
\label{eq53}
\end{equation}
which is the well-known metric of a pulse travelling at speed $v$ with the background speed $c_p$ in the comoving frame. The latter has been used to simulate a black hole, since this is also the Schwarzschild metric in Gullstrand-Painlev\`e coordinates. The experimental design proposed is the same as the one explained above, but with an additional conducting line, where a current pulse with velocity $ v $ is generated, producing a magnetic flux bias, and limited by the propagation velocity of the unbiased SQUIDs, i.e. $ c_ {0} = c (\phi_ {ext} = 0) $  \cite {nation}. Thus, one might think of generating an electromagnetic pulse with the velocity $v$ necessary to generate a CTC. However, the analysis of the null geodesics shows that the negative-time trajectories would require $v>c_p$, which immediately implies $\beta>1/\sqrt{F}$ and thus negative $c_p$ (Figure ~ \ref {fig3}). Thus, we face the same restriction as before, due to the inability of simulating negative speeds of light with this setup. It is worth noting that, due to $ c_{p} $ appears squared in Eq. (\ref{eq53}), we can consider the absolute value of (\ref{eq54}). In this way, we can bypass the problem of simulating a negative light propagation velocity, although we still face the issue of generating a current pulse of negative velocity. Interestingly, the boundary between positive-time and negative-time trajectories is the point $c_{p}^{2}=v^{2}$, which is exactly the condition for the appearance of a horizon in the black-hole interpretation of the metric (\ref{eq53}).
In order to further illuminate the quantum origin of the restrictions preventing us from simulating a CTC, we show in the next section a setup using classical light where the above issues are not present.
\section{\label{level4}Classical simulation of CTCs}
\begin{figure}[!]
\includegraphics[width=0.70\columnwidth]{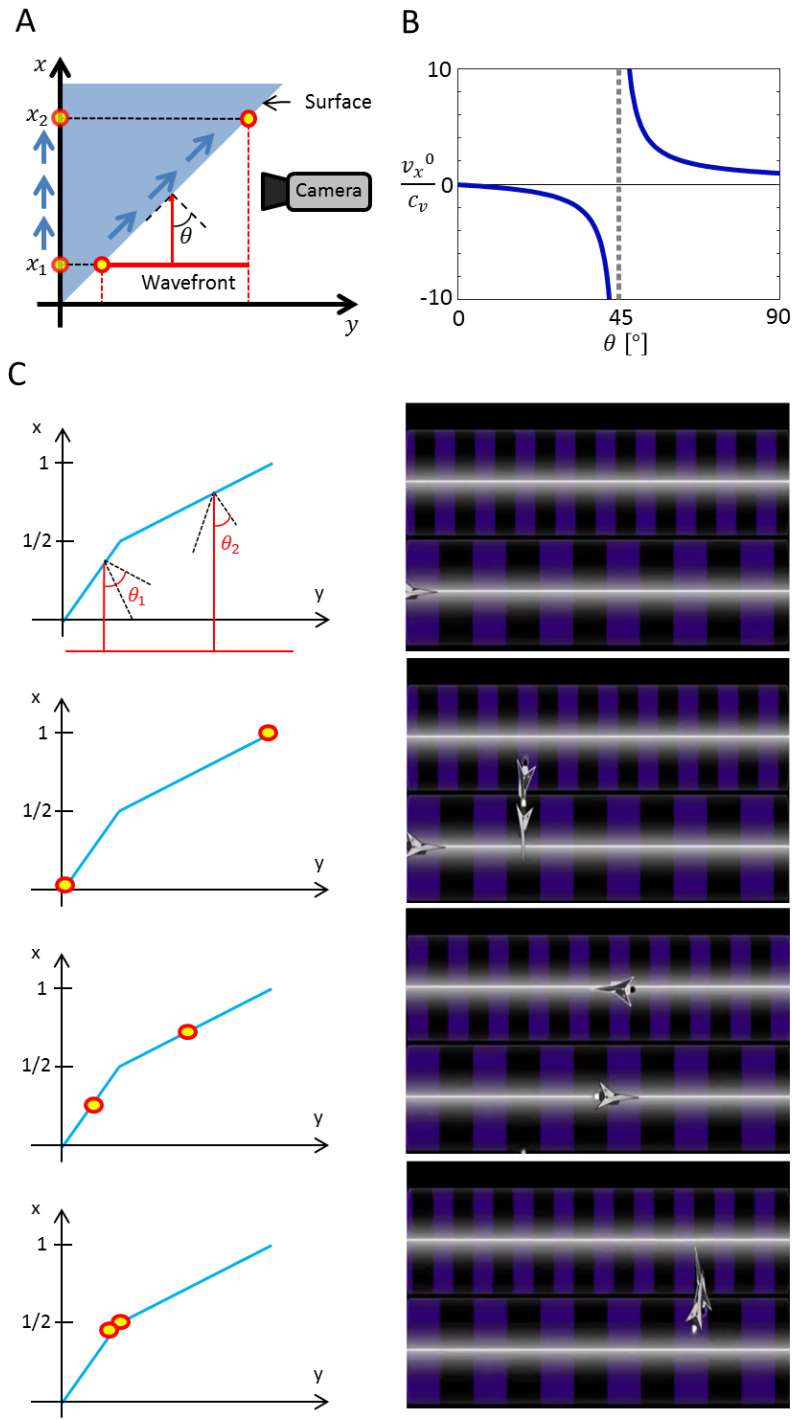}
\caption{\label{fig4} Classical simulation of a CTC. A) Outline of the experimental design \cite {clerici} B) Speed of the scattering source on the x axis observed by the camera for different angles. C) Sequential diagram (from top to bottom) of the simulation of a CTC. (Left) Arrangement of the scattering surfaces and the evolution of the image along the surfaces. In the first scheme, the angle of incidence of the wavefront is made explicit with respect to each surface where the wedges of dashed lines represent an angle of $ 45^{\circ} $ (Right) Diagram of the path of a CTC made by a fictitious rocket obtained from the captures of the facilitated video by \cite {mallary} (https://www.youtube.com/watch?v=ub6PGaygVwA). Note that the first capture does not represent any simulation: the arrival of the rocket cannot be simulated with this setup.}
\end{figure}

Given the impossibility of proposing an effective simulation of a CTC in the quantum system considered, we try to follow the same steps in a classical setup. For this we consider the experiment realized by Clerici et al. \cite {clerici}. This system consists of a light source that emits a wave front that impinges on a surface with an angle $ \theta $, in such a way that the point of intersection of the wave front with the surface moves at a speed $ v = \frac {c_ {v}} {\sin{\theta}} $. This intersection point will be visible due to the scattering of the surface itself, so we will call it the source of scattering. Clearly the scattering source could have superluminal velocities $ v> c_ {v} $. However, this does not pose a problem, because it is not a physical source as such, but a mere cinematical phenomenon \cite {french, gauthier}. Considering the concrete experimental design of Figure ~ \ref {fig4} A, the velocity of the scattering source observed by the camera on the $ x $ axis is given by
\begin{equation}
\ v_{x}^{0}=\frac{c_{v}}{1-\cot \theta }
\label{eq56}
\end{equation}
where $ 0< \theta < \frac{\pi}{4} $ for negative velocities and $ \frac{\pi}{4} < \theta < \frac{\pi}{2} $ for positive velocities (see Figure~\ref{fig4}B), with a singularity at $ \theta = \frac{\pi}{4} $. This behavior resembles that of the effective speed of a photon moving against the direction of motion of a moving wire (\ref{eq41}). Thus, we make the equivalence $ c_{z}^{\beta}=v_{x}^{0} $, and then the incident angle  will be given by:
\begin{equation}
\ \theta =\rm{arccot} \left [ 1- \left ( \frac{1-\sqrt{F}\beta }{\sqrt{F}-\beta } \right )\right ],
\label{eq57}
\end{equation}
which is defined for $ \sim 27^{\circ}< \theta < 90^{\circ} $, when considering the values of $ F $ and $ \beta $. Therefore, it includes the singularity $ \theta=\frac{\pi}{4} $ which corresponds with the negative-time boundary $ \beta=\frac{1}{\sqrt{F}} $. 

The ability of obtaining negative and superluminal light velocities, enables the simulation of a CTC in this setup. For this, two surfaces joined each other with an inclination with respect to the front of incident waves, as can be seen in Figure ~ \ref {fig4} C. Flat surfaces can be used since the speeds of light are always constant. The first surface is arranged in an angle $ \theta_ {1}> 45 ^ {\circ} $ and the second one in an angle $ \theta_ {2} <45 ^ {\circ} $, in such a way that they have positive and negative speeds, respectively. Note that in both cases we have superluminal speeds $ \left | v \right |> c_ {v} $. The first surface is matched with the initial path in the wire at rest, while the second surface represents the subsequent path in the wire with boost. It can be considered that the lengths of the surfaces on the axis $ x $ are normalized to the unit distance, so that the first surface covers $ x \in \left [0, \frac {1} {2} \right] $ and the second surface $ x \in \left [\frac {1} {2}, 1 \right] $. Therefore, the CTC is given by:
\begin{equation}
\ v_{x}^{0}=\frac{c_{v}}{1-\cot \theta },  \cases {\theta_{1} =\rm{arccot}\left [ 1- \frac{1}{\sqrt{F_{1}} } \right ], & $x\in \left [ 0,\frac{1}{2} \right ]$\\ \theta_{2} =\rm{arccot}\left [ 1- \left ( \frac{1-\sqrt{F_{2}}\beta }{\sqrt{F_{2}}-\beta } \right ) \right ], & $x\in \left [\frac{1}{2},1 \right ]$.}\label{eq60}
\end{equation}
where $ \theta_{2} $ has been obtained from (\ref{eq57}) and $ \theta_{1} $ just by making the equivalence $ v_{x}^{0}=c_{z}=c_{v} \sqrt{F_{1}} $. Note that in (\ref{eq60}) it has been considered that the path in the wire at rest and in the moving wire can be carried out at different distances from the central singularities. It suffices simply to take into account the condition (\ref {eq18}) to set the values of $ \theta_ {1} $ and $ \theta_ {2} $, and simulate a CTC. The values of (\ref {eq60}) must be of the same magnitude and opposite sign, to make the path correctly.
In Figure ~ \ref {fig4} C an intuitive scheme of the simulation is represented, where it is compared with the curve proposed by Mallary et al. \cite {mallary} for a fictitious rocket. When the wave front hits the surfaces, two images (scattering sources) are observed at each end that move towards the junction of both surfaces, with the speeds determined by (\ref {eq60}). The image that appears on the left corresponds to the path of the rocket in the wire at rest and the image that appears on the right to the rocket moving in negative time (the rocket appears shaded). Both rockets are at the point where the rocket of the wire at rest passes to the moving wire (in this simulation it would be the point at which the images are annihilated, in the language proposed by Clerici et al. \cite {clerici}). In this case, only one path of a hypothetical infinite loop would have been simulated, the initial arrival of the rocket is not considered.

\section{\label{level5}Conclusions}

We have analyzed possible experimental simulations of CTCs in the space-time recently proposed in \cite{mallary}. Note, that we are not considering a real CTC but a simulation of it, based on an apparent superluminal motion in a flat space-time, that is enough to create a CTC, as Hawking noted \cite{hawking}. We have proposed a classical simulation, based on a recent experiment \cite{clerici} with superluminal optical scattering sources. However, when attempting to propose an analogue quantum simulation by means of an SQUID array, fundamental restrictions appear, preventing us from simulating negative-time trajectories and thus CTCs. This suggests that these restrictions are of quantum origin and therefore they might represent in some way an analogy of the mechanism of chronological protection proposed by Hawking \cite {hawking}. It is worth noting that the analogue of the chronology protection mechanism appears as a technical limitation of the particular analogue setup considered and not as a general feature, as expected in a simulation. 
Paraphrasing Hawking, we might say that it seems that there is a Chronology Protection Agency which prevents the appearance of closed timelike curves and so makes the universe safe for historians even in simulations in analogue systems.

\section*{Acknowledgements}

CS has received financial support through the Postdoctoral Junior Leader Fellowship Programme from “la Caixa” Banking Foundation (code LCF/BQ/LR18/11640005) and from Fundaci\'on General CSIC (ComFuturo Programme).

\end{document}